\documentclass[prd,preprint,
superscriptaddress,showpacs,nofootinbib,%
tightenlines
]{revtex4}
\usepackage{epsfig}
\usepackage{color}
\usepackage{amssymb}

\newcommand{\ben}{\begin{displaymath}}
\newcommand{\een}{\end{displaymath}}
\newcommand{\be}{\begin{equation}}
\newcommand{\ee}{\end{equation}}
\newcommand{\bea}{\begin{eqnarray}}
\newcommand{\eea}{\end{eqnarray}}
\begin{document}

\title{Derivation of spontaneously broken gauge symmetry from  the consistency of effective field theory II:
Scalar field self-interactions and the electromagnetic interaction}
\author{D.~Djukanovic}
 \affiliation{Helmholtz Institute Mainz, University of Mainz, D-55099 Mainz, Germany}
\author{J.~Gegelia}
 \affiliation{Institute for Advanced Simulation, Institut f\"ur Kernphysik
   and J\"ulich Center for Hadron Physics, Forschungszentrum J\"ulich, D-52425 J\"ulich,
Germany}
\affiliation{Tbilisi State  University,  0186 Tbilisi,
 Georgia}
 \author{Ulf-G.~Mei\ss ner}
 \affiliation{Helmholtz Institut f\"ur Strahlen- und Kernphysik and Bethe
   Center for Theoretical Physics, Universit\"at Bonn, D-53115 Bonn, Germany}
 \affiliation{Institute for Advanced Simulation, Institut f\"ur Kernphysik
   and J\"ulich Center for Hadron Physics, Forschungszentrum J\"ulich, D-52425 J\"ulich,
   Germany}
 \affiliation{Tbilisi State  University,  0186 Tbilisi, Georgia}
\date{15. November, 2018}
\begin{abstract}
We extend our  study of deriving the local gauge invariance with spontaneous symmetry breaking 
in the context of an effective field theory by considering self-interactions of the scalar field and
inclusion of the electromagnetic interaction. By analyzing renormalizability and the scale separation
conditions of three-, four- and five-point vertex functions of the scalar field,
we fix the two couplings of the scalar field self-interactions of the leading order Lagrangian.
Next we add the electromagnetic interaction and 
derive conditions relating the magnetic moment of the charged vector boson to its charge and the masses
of the charged and neutral massive vector bosons to each other and the two independent couplings of the theory.
We obtain the bosonic part of the Lagrangian of the electroweak Standard Model
as a unique solution to
the conditions imposed by the self-consistency conditions of the considered effective field theory.

\end{abstract}




\maketitle

\section{Introduction}

Local gauge invariance is taken as an input in the construction of the Standard Model  \cite{Weinberg:mt}. 
On the other hand, a gauge-invariant theory with the spontaneous symmetry breaking can be derived by demanding
tree-order unitarity of the S-matrix \cite{LlewellynSmith:1973ey,Cornwall:1973tb,Cornwall:1974km,Joglekar:1973hh}. 
The modern point of view considers the Standard Model as an 
effective field theory (EFT)  \cite{Weinberg:mt}, in which  tree-order 
unitarity is in any case violated at sufficiently high energies.  This motivated us to address
the issue of deriving the Lagrangian  of the electroweak interaction from the conditions of
self-consistency of EFT. 
In Ref.~\cite{Djukanovic:2018pep} we started  with constructing the  most general Lorentz-invariant
EFT Lagrangian
of three interacting  massive vector bosons and a scalar.  
Non-trivial relations between the coupling constants of the interaction terms of the most
general Lorentz-invariant
Lagrangian of a scalar and vector bosons are imposed by the conditions of consistency with
the second class constraints which must be satisfied by the systems with
spin-one particles \cite{Djukanovic:2010tb}. 
Further restrictions on the interaction terms are imposed by the condition of perturbative renormalizability
in the sense of an EFT and scale separation. The last condition  requires that contributions of higher order
terms of the effective Lagrangian in physical quantities are suppressed by some large scale(s)
(for more details see Ref.~\cite{Djukanovic:2018pep}). To achieve this scale separation we have to demand
that the divergences of loop diagrams contributing in {\it physical} scattering amplitudes generated 
by the leading order Lagrangian should be removable by renormalizing the parameters of the leading
order Lagrangian alone.  

In Ref.~\cite{Djukanovic:2018pep} we considered three- and four-point vertex functions and required
perturbative renormalizability and scale separation. 
This led to conditions imposed on the interaction terms such that  we obtained the Lagrangian 
of spontaneously broken gauge symmetry in the unitary gauge except that the coupling constants of 
the self-interactions of the scalar  field remained unfixed. 
In the current work we analyse one-loop diagrams contributing to three-, four- and five-point
functions of the scalar field and constrain the two free couplings of the 
self-interactions.

Next, in close analogy to Ref.~\cite{Djukanovic:2005ag} we "switch on" the electromagnetic
interaction. By demanding perturbative renormalizability of the obtained effective Lagrangian, we 
relate the magnetic moment of the charged vector boson and  the mixing of the neutral vector bosons
to other parameters of the effective Lagrangian.  This completes the derivation of the bosonic part
of the electroweak Standard Model in the framework of EFT.

\section{Fixing the couplings of the scalar self-interactions}
\label{VMS}

Here, we continue the study of the most  general Lorentz-invariant effective Lagrangian of
a scalar and three massive  vector boson fields respecting  electromagnetic charge conservation,
started in Ref.~\cite{Djukanovic:2018pep}.  Two charged spin-one particles are represented by 
the vector fields $V^\pm_\mu=(V^1_\mu\mp iV^2_\mu)/\sqrt{2}$, while
the third component, $V^3_\mu$, and the scalar field $\Phi$ are charge-neutral. The effective 
Lagrangian contains an infinite number of interaction terms and hence depends on an 
infinite number of parameters. We assume that coupling constants with negative mass dimensions are independent 
from those of positive and zero mass dimensions. In Ref.~\cite{Djukanovic:2018pep} we analyzed the
Lagrangian containing only interaction terms with coupling constants of non-negative dimensions.
By demanding conservation of the second class constraints, perturbative renormalizability in the sense of
EFT and scale separation, we showed that the effective Lagrangian 
can be written in a compact form 
\begin{eqnarray}
{\cal L} & = & -{1\over 4} \ G^a_{\mu\nu} G^{a
\mu\nu} +\frac{1}{2} \,V_\mu^a V^{a \mu}\left( M- \frac{g}{2}\,\Phi \right)^2
-g_{A 1}\epsilon^{abc}\, \epsilon^{\mu\nu\alpha\beta}  V^a_\mu
V^b_\nu
\partial_\alpha V^c_\beta\,, 
\nonumber\\
& + & \frac{1}{2}\, \partial_\mu\Phi \,\partial^\mu\Phi -\frac{m^2}{2}\, \Phi^2
-a\,\Phi -\frac{b}{3!}\,\Phi^3- \frac{\lambda}{4!}\,\Phi^4\,,
\label{EffLagr}
\end{eqnarray}
where
\begin{equation}
G^a_{\mu\nu}=\partial_\mu V^a_\nu - \partial_\nu V^a_\mu -g\,\epsilon^{abc}\,V^b_\mu V^c_\nu\,.
\label{gdefinition}
\end{equation}
The value of the parameter $a$ can be changed by shifting the field $\Phi$ with a constant value.
For convenience  we fix the scalar field such that $a\equiv 0$, i.e. the vacuum expectation value of 
the scalar field is non-vanishing starting at one-loop order.    
The Lagrangian of Eq.~(\ref{gdefinition}) coincides with the SU(2) locally gauge invariant Lagrangian of
scalars and vector  bosons with spontaneous symmetry breaking in the unitary gauge except for the self-interaction 
terms of the scalars. 
As reported in Ref.~\cite{Djukanovic:2018pep} we checked by explicit calculations that no further
constraints on couplings are generated by the 
condition of perturbative renormalizability and scale separation of the three- and four-point
functions of scalar and vector bosons alone. To impose conditions to the two scalar self-interaction
couplings we continue our study by a simultaneous analysis of the three-, four- and five-point functions
of the scalar field.

\begin{figure}[t!]
\epsfig{file=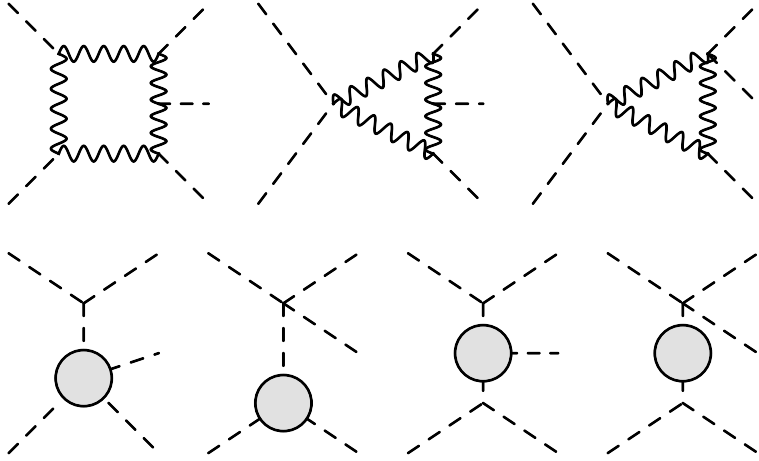, width=10truecm}
\caption[]{\label{SSSSS:fig} One-loop contributions to the 
five-point vertex function of the scalar field. The dashed and wiggly lines correspond to the
scalar and the vector bosons, respectively. Diagrams that are generated via
permutations of external legs  are not shown. Blobs indicate the corresponding
one-loop two-, three- and four-point vertex functions. In the last four diagrams only the
one-particle-irreducible parts are 
taken into account.}
\end{figure}

We impose the on-mass-shell renormalization condition, i.e. require that all divergences in 
physical quantities should be removable by redefining the parameters of the effective Lagrangian.  
As there is no interaction term with five scalar fields in the LO Lagrangian, the sum of divergences
of the one-loop
diagrams contributing to the five-point function should cancel when all external momenta are put on-mass-shell.

The one-loop diagrams which contribute to the five point function of the scalar field are shown in
Fig.~\ref{SSSSS:fig}. 
We apply  dimensional regularization (see, e.g., Ref.~\cite{Collins:1984xc}) and
for calculating the loop diagrams,  we independently use the programs
FeynCalc~\cite{Mertig:1990an,Shtabovenko:2016sxi} and Form~\cite{Kuipers:2012rf}. 
The divergent parts of the one-loop integrals have been checked with the
expressions obtained in  Ref.~\cite{Denner:2005nn}. We have checked the
generation of all diagrams using FeynArts~\cite{Hahn:2000kx}.

Calculating the irreducible one-loop diagrams shown in Fig.~\ref{SSSSS:fig} (plus permutations),
we obtain for the coefficient of the divergent part:
\begin{equation}
\frac{45 i \pi ^2 g^5 m^4}{2 M^5} .
\label{irred5pf}
\end{equation}
The coefficient of the divergent part of the irreducible parts of the reducible diagrams
shown in the second line of 
Fig.~\ref{SSSSS:fig} (plus permutations) has the form
\begin{equation}
\frac{15 i \pi ^2 g^2 \left(3 b^2 g M+9 b g^2 m^2+2 b \lambda 
   M^2+4 g \lambda  m^2 M\right)}{4 M^4} ~.
\label{red5pf}
\end{equation}
Demanding that the sum of Eqs.~(\ref{irred5pf}) and (\ref{red5pf}) vanishes, we obtain 
\begin{equation}
\frac{15 i \pi ^2 g^2 \left(b M+2 g m^2\right) \left(3 b g
   M+3 g^2 m^2+2 \lambda  M^2\right)}{4 M^5}=0,
\label{cond10}
\end{equation}
which has two solutions
\begin{equation}
b=-\frac{2 g m^2}{M},
\label{cond11}
\end{equation}
and
\begin{equation}
\lambda = -\frac{3 g \left(b M+g m^2\right)}{2 M^2} \,.
\label{cond12}
\end{equation}
In the following we will show that only the latter solution leads to a
self-consistent theory.
To that end we substitute the bare parameters of the Lagrangian with
the renormalized ones and the corresponding counterterms
($p=p_R+\sum_{i=1}^\infty \hbar^i\delta p_i$, where $p$ is any of the bare parameters) in
Eqs.~(\ref{cond11}) and (\ref{cond12}) and expand in powers of
$\hbar$. This generates  
the following conditions:
\begin{eqnarray}
\label{bR}
&& b_R =-\frac{2 g_R m_R^2}{M_R},\\
&&
\label{deltab1}
 \text{$\delta $b}_1= - \frac{2 m_R M_R \left(2 \text{$\delta $m}_1
   g_R+\text{$\delta $g}_1 m_R\right)-2 \text{$\delta $M}_1
   g_R m_R^2}{M_R^2},\\
&&    \cdots \nonumber ,
\end{eqnarray}
and 
\begin{eqnarray}
&& \lambda_R = -\frac{3 g_R (g_R m_R^2 + b_R M_R)}{2 M_R^2},\\
&&
\label{deltalambda1}
 \delta
   \lambda_1 = - \frac{1}{2 M_R^3} \biggl[ 3 g_R M_R \left(-\text{$\delta $M}_1
   b_R+2 \text{$\delta $g}_1 m_R^2+\text{$\delta $b}_1
   M_R\right)+ 3 M_R^2  \text{$\delta $g}_1 b_R  \nonumber\\  && \qquad \ \ \ +6 g_R^2 m_R \left(\text{$\delta $m}_1
   M_R-\text{$\delta $M}_1 m_R\right) \biggr] ,\\
   && \cdots \nonumber , 
\label{cond22}
\end{eqnarray}   
respectively.
Equations~(\ref{deltab1}) and (\ref{deltalambda1}) impose conditions on the divergent parts
of one-loop vertex functions, the 
divergent parts of which are cancelled by the corresponding one-loop counterterms 
($\delta b _1$, $\delta \lambda _1$, $\text{$\delta $g}_1$, $\text{$\delta $M}_1$, $\text{$\delta $m}_1$). 

\begin{figure}[t!]
\epsfig{file=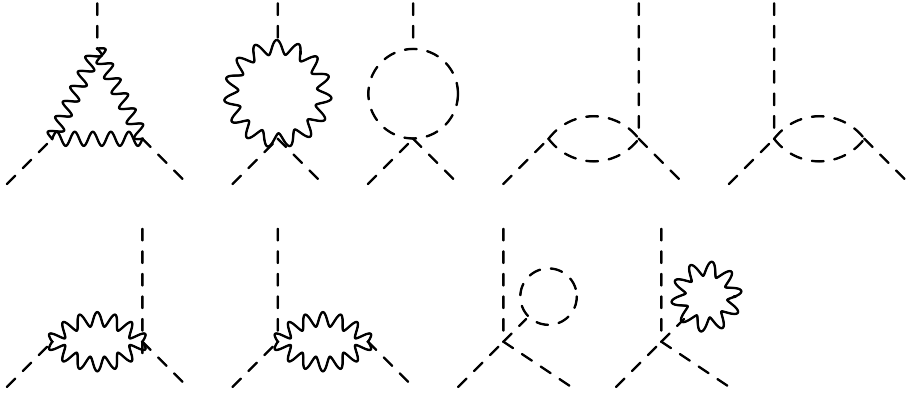, width=12truecm}
\caption[]{\label{SSS:fig} One-loop contributions to the
three-scalar vertex function. The dashed and the wiggly lines correspond to the
scalar and the vector bosons, respectively. }
\end{figure}

\begin{figure}[t!]
\epsfig{file=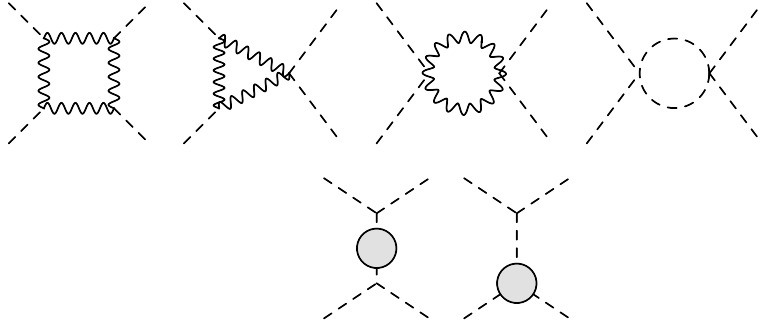, width=10truecm}
\caption[]{\label{SSSS:fig} One-loop contributions to the
four-scalar vertex function. The dashed and wiggly lines correspond to the
scalar and the vector bosons, respectively. Diagrams that can be
generated via permutations of external legs are not shown. Blobs indicate 
the corresponding one-loop two- and 
three-point vertex functions. In last two diagrams only the one-particle-irreducible parts are 
taken into account.}
\end{figure}

From the calculations of Ref.~\cite{Djukanovic:2018pep} we have
\begin{eqnarray}
\text{$\delta $g}_1 &=& -\frac{43}{6} \pi ^2 g_R^3\,,\nonumber\\
\text{$\delta $M}_1 &=&  \frac{\pi ^2 g_R \left(6 b_R m_R^2-g_R M_R \left(59 m_R^2+54
   M_R^2\right)\right)}{12 m_R^2}\,,\nonumber\\
\text{$\delta $m}_1 &=& \frac{\pi ^2 \left(36 b_R g_R M_R^5+3 g_R^2 \left(-6 m_R^4
   M_R^2+18 m_R^2 M_R^4+m_R^6\right)+4 m_R^4 M_R^2 \lambda
   _R\right)}{8 m_R^3 M_R^2}\,.
\label{oldC}
\end{eqnarray}
The one-loop counterterm $\delta g_1$ was obtained by demanding that the three-point vertex function of the 
renormalized vector fields for all three external momenta taken on mass-shell is finite. 
The same expression takes also care of the cancellation of the divergences in the four-point vertex
function of the
vector fields when all external momenta are taken on mass shell. The counterterms $\delta M_1$ and
$\delta m_1$ were determined from the condition that the pole masses of the vector bosons and the scalar
particle are finite at one-loop order.  

The counterterms $\text{$\delta $b}_1$ and $\delta \lambda _1$ are obtained by calculating the divergent parts of 
the one-loop diagrams shown in Figs.~\ref{SSS:fig} and \ref{SSSS:fig}, respectively:
\begin{eqnarray}
\delta \lambda _1 &=&
\frac{3 \pi ^2}{4 M_R^4} \biggl(g_R^2 M_R^2 \left(3 b_R^2+4 \lambda _R
   \left(m_R^2-3 M_R^2\right)\right)+12 b_R g_R^3 m_R^2 M_R\nonumber\\
   &+& 9  g_R^4 \left(m_R^4+M_R^4\right)+4 M_R^4 \lambda
   _R^2\biggr)\,,\label{uc1}\\
\text{$\delta $b}_1 & = & \frac{\pi ^2 }{4 m_R^2 M_R^3} \biggl( 9 b_R g_R^2 m_R^2 M_R \left(m_R^2-3
   M_R^2\right)+8 b_R m_R^2 M_R^3 \lambda _R \nonumber\\
   &+& 9 g_R^3
   \left(m_R^6-6 m_R^2 M_R^4\right)+36 g_R M_R^6 \lambda
   _R\biggr)\,.
\label{newC}
\end{eqnarray}
By taking into account Eqs.~(\ref{oldC}) and (\ref{newC})  in Eq.~(\ref{deltab1}) and using Eq.~(\ref{bR}),
we obtain the following condition which has to be satisfied by renormalized 
parameters $\lambda_R$, $g_R$, $m_R$ and $M_R$:
\begin{equation}
\frac{g_R \left(g_R^2 \left(5 m_R^6-54 m_R^2
   M_R^4\right)-8 m_R^4 M_R^2 \lambda _R+36 M_R^6 \lambda
   _R\right)}{m_R^2 M_R^3}=0.
   \label{eqlambda}
\end{equation}
The solution to Eq.~(\ref{eqlambda}) 
\begin{equation}
\lambda _R = \frac{g_R^2 \left(54 m_R^2
   M_R^4-5 m_R^6\right)}{36 M_R^6-8 m_R^4
   M_R^2},
\label{lambdarelbare}
\end{equation}
does not lead to a self-consistent condition. This is because a relation like
Eq.~(\ref{lambdarelbare}) can be satisfied for an
arbitrary renormalization scheme only if the corresponding 
bare couplings satisfy the same condition. This, however, imposes 
the following condition on  the counterterms
\begin{eqnarray}
 \delta
   \lambda_1& = &\frac{g_R m_R }{2 M_R^3 \left(2 m_R^4-9 M_R^4\right){}^2} 
   \biggl[  10 m_R^9 \left(\text{$\delta $g}_1 M_R-\text{$\delta
   $M}_1 g_R\right)-9 m_R^5 M_R^4 \left(17 \text{$\delta $g}_1 M_R-3
   \text{$\delta $M}_1 g_R\right) \nonumber\\
   &+& 486 m_R M_R^8 \left(\text{$\delta $g}_1
   M_R-\text{$\delta $M}_1 g_R\right)+10 \text{$\delta $m}_1 g_R m_R^8
   M_R  
   - 27 \text{$\delta $m}_1 g_R m_R^4 M_R^5+486 \text{$\delta $m}_1 g_R
   M_R^9\biggr] ,
   \nonumber
\label{condcc}
\end{eqnarray}
which is not satisfied by the expressions specified in Eqs.~(\ref{oldC}) and (\ref{uc1}). 
That is, the first solution to the condition of Eq.~(\ref{cond10}), given in Eq.~(\ref{cond11}),
does not lead to self-consistent renormalization.

Next, by taking into account Eqs.~(\ref{oldC},\ref{newC})  in Eq.~(\ref{deltalambda1}),  we
obtain the following condition which has to be satisfied by renormalized parameters $b_R$, $g_R$, $m_R$ and $M_R$:
\begin{eqnarray}
&& \frac{g_R^2 \left(7 b_R g_R m_R^2 M_R+2
   b_R^2 M_R^2+6 g_R^2 m_R^4\right)}{M_R^4} =0\,.
\label{condbR}
\end{eqnarray}
Eq.~(\ref{condbR}) has two solutions
\begin{equation}
b_R = - \frac{2 g_R m_R^2}{M_R}, \ \ \ b_R = - \frac{3 g_R m_R^2}{2 M_R}\,.
\label{bsol}
\end{equation}
However, analogously to the above case, only one, namely
\begin{equation}
b_R = - \frac{3 g_R m_R^2}{2 M_R}
\label{bsol}
\end{equation}
leads to a self-consistent condition. 
Substituting Eq.~(\ref{bsol}) in Eq.~(\ref{lambdarelbare}) leads to
\begin{equation}
\lambda_R = \frac{3 g_R^2 m_R^2}{4 M_R^2}\,.
\label{lR}
\end{equation}
As mentioned above, Eqs.~(\ref{bsol}) and (\ref{lR}) can be satisfied only if analogous
relations hold for corresponding bare quantities, i.e. we have
\begin{equation}
b = - \frac{3 g m^2}{2 M}\,,  \ \ \ \lambda = \frac{3 g^2 m^2}{4 M^2}\,.
\label{bl}
\end{equation}
This fixes uniquely the coupling constants of the scalar self-interactions to the values corresponding to 
spontaneously broken gauge symmetry in the unitary gauge.

\section{Inclusion of the electromagnetic interaction}
To have massless spin-one particles, the photons, in the spectrum of the theory it is necessary that
the Lagrangian is invariant under local gauge $U(1)$ transformations \cite{Weinberg:mt}.  
Therefore,  we introduce an Abelian gauge field $B_\mu$ and its coupling to the charged vector fields and also a 
gauge-invariant mixing term of the neutral vector fields.  
The resulting Lagrangian reads 
(with  $a=0$)
\begin{eqnarray}
{\cal L} & = & -{1\over 4} \, B_{\mu\nu} B^{\mu\nu} -{1\over 4} \, G^a_{\mu\nu} G^{a
\mu\nu} +\frac{1}{2} \,V_\mu^a V^{a \mu}\left( M- \frac{g}{2}\,\Phi \right)^2
-g_{A 1}\epsilon^{abc}\, \epsilon^{\mu\nu\alpha\beta}  V^a_\mu
V^b_\nu
\partial_\alpha V^c_\beta\,, 
\nonumber\\
& + & \frac{c}{2}\, B^{\mu\nu}\, V^3_{\mu\nu} +\frac{\kappa}{2}\,\epsilon^{3ab}\,B^{\mu\nu} V^a_\mu V^b_\nu  
 +  \frac{1}{2}\, \partial_\mu\Phi \,\partial^\mu\Phi -\frac{m^2}{2}\, \Phi^2 \left( 1 - \frac{g}{4 M}\,\Phi \right)^2\,,
\label{EffLagrEM}
\end{eqnarray}
where
\begin{eqnarray}
B_{\mu\nu} & = & \partial_\mu B_\nu - \partial_\nu B_\mu\,, \ \ \ V^a_{\mu\nu} = \partial_\mu V^a_\nu - \partial_\nu V^a_\mu\, ,\nonumber\\
G^a_{\mu\nu} & = & V^a_{\mu\nu}-g\,\epsilon^{abc}\,V^b_\mu V^c_\nu + e\,\epsilon^{3ab}\,\left(B_\mu V^b_\nu -B_\nu V^b_\mu \right) ,
\label{gdefinitionEM}
\end{eqnarray}
and we have substituted the expressions of Eq.~(\ref{bl}).
To diagonalize the Lagrangian, Eq.~(\ref{EffLagrEM}), we introduce new vector fields $A_\mu$, $Z_\mu$ 
and $W_\mu^\pm$ as follows: 
\begin{eqnarray}
B_\mu & = & A_\mu + \frac{c }{\sqrt{1-c^2}} \,Z_\mu \,,\nonumber\\
V^\pm_\mu &=& (V^1_\mu\mp iV^2_\mu)/\sqrt{2} =W^\pm_\mu , \ V^3_\mu = \frac{Z_\mu}{\sqrt{1-c^2}}\,.
\label{newfields}
\end{eqnarray}

Next, we  analyze the conditions of perturbative renormalizability. We again
split the bare parameters as $p=p_R+\sum_{i=1}^\infty \hbar^i\delta p_i$ and
fix the counterterms such that they cancel divergences in physical quantities. 

The undressed propagators of the $Z$ and $W^\pm$ vector bosons read
\begin{equation}
i\,S^{Z,W}_{0,\mu\nu}(p) = -i
\,\frac{g_{\mu\nu}-\frac{p_\mu p_\nu}{M_{Z,W}^2}}{p^2 -
M_{Z,W}^2+i\,0^+}\,,\label{bareprop}
\end{equation}
where $M_W=M$ and $M_Z=M/\sqrt{1-c^2}$.
We parameterize the sum of all one-particle-irreducible diagrams
contributing to the two-point functions as
\begin{equation}
i\,\Pi^{Z,W}_{\mu\nu}(p)=i \,\left[ g_{\mu\nu}\Pi^{Z,W}_{1}
(p^2)+p_\mu p_\nu\,\Pi^{Z,W}_2 (p^2)\right]\,. \label{VSEpar}
\end{equation}
The corresponding dressed propagators have 
the form
\begin{equation}
i\, S^{Z,W}_{\mu\nu}(p) = -i
\,\frac{g_{\mu\nu}-p_\mu p_\nu\frac{1+
\Pi^{Z,W}_2(p^2)}{M_{Z,W}^2+\Pi^{Z,W}_1(p^2)+p^2 \Pi^{Z,W}_2(p^2)}}{p^2 -
M_{Z,W}^2-\Pi^{Z,W}_1(p^2)+i\,0^+}\,.\label{dressedprop}
\end{equation}
The pole masses are obtained from the solutions to
the following equations:
\begin{equation}
z_{Z,W} - M_{Z,W}^2-\Pi_1(z_{Z,W})=0\,. \label{poleequation1}
\end{equation}
In the vicinity of the pole the dressed propagators can be expanded as
\begin{equation}
i\, S^{Z,W}_{\mu\nu}(p) = -i \left[\frac{Z^r_{Z,W}
\left(g_{\mu\nu}-\frac{p_\mu p_\nu}{z_{Z,W}}\right)}{p^2
-z_{Z,W}+i\,0^+}+R\right]\,,\label{dressedpropnearpole}
\end{equation}
where
$$
Z^r_{Z,W}=\frac{1}{1- \Pi^{Z,W\,'}_1(z_{Z,W})} 
$$
is the wave-function renormalization constant 
and $R$ denotes the non-pole part.

From Eq.~(\ref{poleequation1}) at one-loop order we have 
\begin{equation}
z_{Z,W} = M_{Z,W}^2+\Pi_1(M_{Z,W}^2)\,. \label{poleequation1sol}
\end{equation}
Substituting in Eq.~(\ref{poleequation1sol}) 
$$M_W=M=M_R+\hbar \, \delta M_1$$ 
and 
$$M_Z=\frac{M}{\sqrt{1-c^2}}=\frac{M_R}{\sqrt{1-c_R^2}}-\frac{\hbar \left(\delta M_1
   \left(c_R^2-1\right)- \delta c_1 c_R
   M_R\right)}{\left(1-c_R^2\right){}^{3/2}}
$$
and demanding that the pole masses of both 
the $Z$ and the $W$ bosons  must be finite quantities, we obtain
\begin{eqnarray}
\delta M_1 &=& \frac{\pi ^2 }{12
   \left(c_R^2-1\right){}^3 m_R^2} 
   \Biggl\{ M_R \biggl[ 4 \left(c_R^2-1\right) e_R
   m_R^2 \left(28 c_R^3 g_R-25 c_R g_R-3 c_R^2 \kappa _R-6
   \kappa _R\right)\nonumber\\
   &+& 4 \left(17 c_R^4-28 c_R^2+11\right)
   e_R^2 m_R^2+g_R^2 \biggl(
   \left(37 c_R^6-32 c_R^4-73
   c_R^2+59\right) m_R^2 \nonumber\\
   &-& 18 \left(2 c_R^6-6 c_R^4+7
   c_R^2-3\right) M_R^2\biggr)
   +2 c_R \left(23 c_R^4-78
   c_R^2+46\right) g_R m_R^2 \kappa _R\nonumber\\
   &+& \left(-3 c_R^4-17
   c_R^2+11\right) m_R^2 \kappa _R^2\biggr]-\frac{9
   \left(c_R^2-1\right){}^3 g_R^2 m_R^4}{M_R}\Biggr\}\,,
   \nonumber\\
\delta c_1 &=& -\frac{\pi ^2}{12 c_R
   \left(c_R^2-1\right) ^2}  
   \biggl[ 8 \left(c_R^2-1\right) e_R \left(-2
   c_R^3 g_R+3 c_R g_R+9 c_R^4 \kappa _R-6 c_R^2 \kappa
   _R+3 \kappa _R\right)\nonumber\\
   &+& 4 \left(3 c_R^6-22 c_R^4+30
   c_R^2-11\right) e_R^2+26 c_R^5 g_R \kappa _R+c_R^4
   \left(106 g_R^2+\kappa _R^2\right)  \nonumber\\
   &+& 20 c_R^3 g_R \kappa
   _R+c_R^2 \left(18 \kappa _R^2-61 g_R^2\right)-30 c_R g_R
   \kappa _R-37 c_R^6 g_R^2-11 \kappa _R^2
   \biggr]\,.
\label{deltaMandc}
\end{eqnarray}
Next,  we calculate the one-loop contributions to the $\Phi ZZ$ and $\Phi W^+W^-$ vertex functions
and demand that the divergences are cancelled in both quantities when taken on mass-shell.
Using Eq.~(\ref{deltaMandc}), we obtain two expressions of $\delta g_1$ resulting
from the conditions of the finiteness of the $\Phi ZZ$ and $\Phi W^+W^-$ vertex functions:
\begin{eqnarray}
\delta g_1^{(Z)} &=& \frac{\pi ^2 g_R }{12
   \left(c_R^2-1\right){}^3 M_R^4}
\biggl[
c_R^6 \left(8 e_R m_R^2 M_R^2 \kappa
   _R+e_R^2 \left(8 m_R^2 M_R^2-18 m_R^4\right) 
   + g_R^2
   \left(6 m_R^2 M_R^2  \right. \right.  \nonumber\\
  &+& \left. \left. 19 M_R^4\right)+2 m_R^2 \kappa _R^2
   \left(9 m_R^2-8 M_R^2\right)\right)+2 c_R^5 g_R
   \left(e_R \left(12 m_R^2 M_R^2-18 m_R^4+56
   M_R^4\right) \right.\nonumber\\
   &+& \left. \kappa _R \left(-12 m_R^2 M_R^2+18 m_R^4+23
   M_R^4\right)\right)+c_R^4 \left(-4 e_R M_R^2 \kappa _R
   \left(4 m_R^2+3 M_R^2\right) \right. \nonumber\\
   &+& \left.  2 e_R^2 \left(-8 m_R^2
   M_R^2+9 m_R^4+34 M_R^4\right)+g_R^2 \left(31 M_R^4-18
   m_R^2 M_R^2\right)+\kappa _R^2 \left(26 m_R^2 M_R^2  \right. \right. \nonumber\\
  & - & \left.\left. 18
   m_R^4-3 M_R^4\right)\right)+4 c_R^3 g_R \left(e_R
   \left(-12 m_R^2 M_R^2+9 m_R^4-53 M_R^4\right)-3 \kappa
   _R \left(-3 m_R^2 M_R^2  \right.\right. \nonumber\\ 
  & + & \left. \left. 3 m_R^4+13
   M_R^4\right)\right)-c_R^2 \left(4 e_R M_R^2 \kappa _R
   \left(3 M_R^2-2 m_R^2\right)+2 e_R^2 \left(-4 m_R^2
   M_R^2+3 m_R^4+56 M_R^4\right)\right.
   \nonumber\\
   &+& \left. g_R^2 \left(145 M_R^4-12
   m_R^2 M_R^2\right)+\kappa _R^2 \left(10 m_R^2 M_R^2-6
   m_R^4+17 M_R^4\right)\right)
    \nonumber\\
   &-&  4 c_R g_R \left(e_R
   \left(-6 m_R^2 M_R^2+3 m_R^4-25 M_R^4\right)+\kappa _R
   \left(3 m_R^2 M_R^2-3 m_R^4-23 M_R^4\right)\right)
   \nonumber\\
  & + & 12
   c_R^7 g_R m_R^4 \left(e_R-\kappa _R\right)+6 c_R^8 m_R^4
   \left(e_R^2-\kappa _R^2\right)+M_R^4 \left(24 e_R \kappa
   _R+44 e_R^2+86 g_R^2+11 \kappa _R^2\right)\biggr]\,,\nonumber\\
   \delta g_1^{(W)} &=& \frac{\pi ^2 g_R}{12 \left(c_R^2-1\right){}^3
   M_R^4} 
   \biggl[ 2 \left(c_R^2-1\right) e_R
   \left(c_R^3 g_R \left(3 m_R^4+56 M_R^4\right)-c_R g_R
   \left(3 m_R^4+50 M_R^4\right) \right. \nonumber\\
&  -& \left. 2 c_R^2 M_R^2 \kappa _R
   \left(2 m_R^2+3 M_R^2\right)+4 M_R^2 \kappa _R
   \left(m_R^2-3 M_R^2\right)\right)+\left(c_R^2-1\right)
   e_R^2 \left(4 \left(c_R^2-1\right) m_R^2 M_R^2 \right. \nonumber\\
&   + & \left. 3
   \left(c_R^2-1\right) m_R^4+4 \left(17 c_R^2-11\right)
   M_R^4\right)+2 c_R^5 g_R \kappa _R \left(-20 m_R^2
   M_R^2+2 m_R^4+23 M_R^4\right)
    \nonumber\\
   &+& c_R^4 \left(g_R^2 \left(58
   m_R^2 M_R^2-10 m_R^4+31 M_R^4\right)+\kappa _R^2
   \left(-10 m_R^2 M_R^2+2 m_R^4-3 M_R^4\right)\right)
    \nonumber\\
   &-& 4
   c_R^3 g_R \kappa _R \left(-23 m_R^2 M_R^2+2 m_R^4+39
   M_R^4\right)+c_R^2 \left(g_R^2 \left(-32 m_R^2 M_R^2+5
   m_R^4-145 M_R^4\right) 
   \right. \nonumber\\
  & + & \left. \kappa _R^2 \left(26 m_R^2
   M_R^2-4 m_R^4-17 M_R^4\right)\right)+4 c_R g_R \kappa _R
   \left(-13 m_R^2 M_R^2+m_R^4+23 M_R^4\right)
    \nonumber\\
 &  + & c_R^6 g_R^2
   \left(-26 m_R^2 M_R^2+5 m_R^4+19 M_R^4\right)+86 g_R^2
   M_R^4-16 m_R^2 M_R^2 \kappa _R^2  \nonumber\\ 
   &+&2 m_R^4 \kappa _R^2+11
   M_R^4 \kappa _R^2\biggr]\,.
\label{twoconditions}
\end{eqnarray}
The two expressions for the same counterterm have to coincide, leading to the following condition:
 \begin{eqnarray}
 && \frac{\pi ^2 g_R m_R^2}{12 \left(c_R^2-1\right){}^2 M_R^4} 
 \biggl[2 \left(c_R^2-1\right) e_R
   \left(3 c_R g_R \left(4 M_R^2-3 m_R^2\right)+6 c_R^3 g_R
   m_R^2+4 c_R^2 M_R^2 \kappa _R+4 M_R^2 \kappa
   _R\right) \nonumber\\
&& + \left(c_R^2-1\right) e_R^2 \left(\left(6
   c_R^4-6 c_R^2-3\right) m_R^2+4 \left(2 c_R^2-1\right)
   M_R^2\right)
   +c_R^4 \left(g_R^2 \left(32 M_R^2-5
   m_R^2\right) \right.  \nonumber\\
 &&  \left.+4 \kappa _R^2 \left(3 m_R^2-4
   M_R^2\right)\right)+4 c_R^3 g_R \kappa _R \left(5
   m_R^2+4 M_R^2\right)+c_R^2 \left(g_R^2 \left(5 m_R^2-44
   M_R^2\right)
 \right.  \nonumber\\
 &&  \left.   
   +4 \kappa _R^2 \left(5 M_R^2-2
   m_R^2\right)\right)-8 c_R g_R \kappa _R \left(m_R^2+5
   M_R^2\right)
   -12 c_R^5 g_R m_R^2 \kappa _R \nonumber\\
   && -6 c_R^6 m_R^2
   \kappa _R^2+2 \kappa _R^2 \left(m_R^2-8
   M_R^2\right)\biggr]=0\,.
  \label{condition1}
  \end{eqnarray}
By demanding that the same counterterm $\delta g_1$ in combination with $\delta M_1$ and 
\begin{equation}
\delta m_1 = \frac{3 \pi ^2 g_R^2 m_R \left(\left(c_R^2-1\right)
   m_R^2+\left(3-2 c_R^2\right) M_R^2\right)}{4
   \left(c_R^2-1\right) M_R^2}\,,
   \label{deltam1}
   \end{equation}
removes the divergences from the $\Phi\Phi\Phi$ vertex function, we obtain another condition:
 \begin{eqnarray}  
&&   \frac{\pi ^2 c_R g_R m_R^4 }{4
   \left(c_R^2-1\right){}^2 M_R^5}
  \biggl[ c_R^3 \left(e_R^2 \left(4
   M_R^2-6 m_R^2\right)+4 e_R M_R^2 \kappa _R+3 g_R^2
   M_R^2+2 \kappa _R^2 \left(3 m_R^2-4
   M_R^2\right)\right)  
\nonumber\\   
&&  -12 c_R^2 g_R \left(e_R-\kappa
   _R\right) \left(m_R^2-M_R^2\right)+c_R \left(e_R^2
   \left(3 m_R^2-4 M_R^2\right)-4 e_R M_R^2 \kappa _R-6
   g_R^2 M_R^2
\right. 
\nonumber\\   
  && \left.   
   +\kappa _R^2 \left(5 M_R^2-3
   m_R^2\right)\right)+6 c_R^4 g_R m_R^2 \left(e_R-\kappa
   _R\right)+3 c_R^5 m_R^2 \left(e_R^2-\kappa _R^2\right)
\nonumber\\   
  && 
+6
   g_R \left(e_R \left(m_R^2-2 M_R^2\right)
   +\kappa _R
   \left(M_R^2-m_R^2\right)\right)\biggr]
=0\,.
  \label{condition2}
  \end{eqnarray}

One more condition is obtained by calculating the one-loop contributions to the $Z W^{+} W^{-} $ vertex
function and demanding that its divergent 
part proportional to the Lorentz-structure with the product of three momenta (i.e. containing no metric
tensor) vanishes.  The resulting expression has the form:
\begin{eqnarray}
\frac{c_R (e_R-\kappa_R )^2 (4 c_R g_R+3 \,e_R+\kappa_R
   )}{\left(1-c_R^2\right)^{3/2} M_R^2} =0\,.
   \label{onmorecond}
   \end{eqnarray}
Eqs.~(\ref{condition1}), (\ref{condition2}) and (\ref{onmorecond}) fix $\kappa_R$ and $c_R$ to the
following unique  expressions:
\begin{equation}
\kappa_R=e_R, \  \  c_R = -\frac{e_R}{g_R}\,.
\label{solkappaandc}
\end{equation}
As argued above, Eq.~(\ref{solkappaandc}) leads to analogous relations for corresponding bare
parameters of the effective Lagrangian.
  
Thus all parameters of our LO effective Lagrangian of interacting photons, a scalar and massive neutral
and two charged vector bosons are uniquely fixed by the self-consitsency conditions such that the 
Lagrangian corresponding to the  electroweak sector of the  Standard Model in unitary gauge is obtained.

\section{Summary}
\label{Concl}

In the current work we extended the study of Ref.~\cite{Djukanovic:2018pep} where  
following the modern point of view of the Standard Model as the leading order approximation of 
an effective field theory we analyzed the most general Lorentz-invariant leading order 
effective Lagrangian of massive vector bosons interacting with a massive scalar field. Here
leading order means interaction terms with couplings of non-negative mass dimensions.

In Ref.~\cite{Djukanovic:2018pep} we analyzed the conditions of perturbative renormalizability
and scale separation applied to all  three- and four-point functions at one-loop order.  
These conditions in combination with the second class constraints imposed on systems with spin-one particles 
led to severe restrictions on the interaction terms of the leading order Lagrangian. However,
two coupling constants of the self-interactions of the scalar field remained unfixed. 
In the current work, using the condition of perturbative  renormalizability and
scale separation for three-, four- and five-point functions of the scalar field
at one-loop order, we were able to fix the remaining two free couplings.  Next,
we included the coupling to the electromagnetic interaction and again demanded
self-consistency in the sense of EFT.  Analyzing the  renormalizability
conditions of the various three-point vertex functions 
we fixed the two additional free parameters appearing in the most general effective Lagrangian.
As the result of this analysis the Lagrangian 
with spontaneously broken SU(2)$\times$U(1) gauge symmetry taken in unitary
gauge naturally appears as the unique leading-order Lagrangian of a self-consistent EFT
of a massive scalar interacting with neutral and charged massive vector bosons
and the electromagnetic field.  
It is well-known that such a Lagrangian leads to a well-defined finite $S$-matrix \cite{'tHooft:1971fh}. 

The inclusion of the fermionic degrees of freedom is the last step 
for completing this program  of deriving the leading order EFT Lagrangian of the electroweak interaction
and will be considered in a forthcoming publication.

\acknowledgments
This work was supported in part  by the DFG and NSFC through funds provided to the Sino-German
CRC 110 ``Symmetries and the Emergence of Structure in QCD'' (NSFC Grant No.  11621131001, DFG
Grant  No.  TRR110),  by  the  VolkswagenStiftung (Grant No.   93562), by 
the CAS President's International Fellowship Initiative (PIFI) (Grant No. 2018DM0034) and
by the Georgian Shota Rustaveli National Science Foundation (Grant No. FR17-354).


\end{document}